\documentclass[11pt,epsf]{article}
 \usepackage{amsmath}
 \usepackage{graphicx}
 \usepackage[merge,numbers,compress]{natbib}
 \usepackage[T1]{fontenc}
 \usepackage{booktabs}
 \usepackage{xcolor} 
 \usepackage{xspace}
 \usepackage{dcolumn}
 \usepackage{placeins}
 \usepackage{amssymb}
 \usepackage[colorlinks=true,citecolor=blue!50!black,linkcolor=black]{hyperref}
 \usepackage{caption}
 \usepackage{soul}
 \usepackage
 [subrefformat=parens,position=top,skip=-15pt,margin=15pt,justification=justified,singlelinecheck=false]
 {subcaption}
 \usepackage{array,multirow}

\newcommand{\rd}{\mathrm d}

\def\be{\begin{equation}}
\def\ee{\end{equation}}

\newcommand{\Pe}{\ensuremath{\text{e}}\xspace}

\newcommand{\PW}{\ensuremath{\text{W}}\xspace}
\newcommand{\PZ}{\ensuremath{\text{Z}}\xspace}

\newcommand{\MWOS}{\ensuremath{M_\PW^\text{OS}}\xspace}
\newcommand{\MW}{\ensuremath{M_\PW}\xspace}
\newcommand{\MZOS}{\ensuremath{M_\PZ^\text{OS}}\xspace}
\newcommand{\MZ}{\ensuremath{M_\PZ}\xspace}

\newcommand{\GZOS}{\ensuremath{\Gamma_\PZ^\text{OS}}\xspace}

\newcommand{\GWOS}{\ensuremath{\Gamma_\PW^\text{OS}}\xspace}

\newcommand{\GeV}{\ensuremath{\,\text{GeV}}\xspace}

\newcommand{\GF}{\ensuremath{G_\mu}}

\newcommand{\MVOS}{\ensuremath{M_{\text{V}}^\text{OS}}\xspace}%
\newcommand{\GVOS}{\ensuremath{\Gamma_{\text{V}}^\text{OS}}\xspace}%

\newcommand{\newc}{\newcommand}
\newc{\bi}{\begin{itemize}}
\newc{\ei}{\end{itemize}}
\newc{\benu}{\begin{enumerate}}
\newc{\eenu}{\end{enumerate}}
\newc{\bc}{\begin{center}}
\newc{\ec}{\end{center}}
\newc{\bfig}{\begin{figure}}
\newc{\efig}{\end{figure}}
\newc{\qbar}{\bar{q}}
\newc{\go}{\tilde{g}}
\newc{\PB}{\textsc{Powheg-Box}}

\newcolumntype{.}{D{.}{.}{-1}}
\newcolumntype{d}[1]{D{.}{.}{#1}}

\colorlet{tableoverheadcolor}{gray!37.5}
\colorlet{tableheadcolor}{gray!25}
\colorlet{tablerowcolor}{gray!12.5}

\newlength{\width}
\newlength{\height}

\def\draftdate{\relax}
\def\mda{\relax}
\def\mua{\relax}
\def\mla{\relax}
\def\draft{
\def\thtystars{******************************}
\def\sixtystars{\thtystars\thtystars}
\typeout{}
\typeout{\sixtystars**}
\typeout{* Draft mode!
         For final version remove \protect\draft\space in source file *}
\typeout{\sixtystars**}
\typeout{}
\def\draftdate{\today}
\def\mua{\marginpar[\boldmath\hfil$\uparrow$]%
                   {\boldmath$\uparrow$\hfil}\color{black}%
                    \typeout{marginpar: $\uparrow$}\ignorespaces}
\def\mda{\color{red}\marginpar[\boldmath\hfil$\downarrow$]%
                   {\boldmath$\downarrow$\hfil}%
                    \typeout{marginpar: $\downarrow$}\ignorespaces}
\def\mla{\marginpar[\boldmath\hfil$\rightarrow$]%
                   {\boldmath$\leftarrow $\hfil}%
                    \typeout{marginpar: $\leftrightarrow$}\ignorespaces}
\def\Mua{\marginpar[\boldmath\hfil$\Uparrow$]%
                   {\boldmath$\Uparrow$\hfil}\color{black}%
                    \typeout{marginpar: $\uparrow$}\ignorespaces}
\def\Mda{\color{red}\marginpar[\boldmath\hfil$\Downarrow$]%
                   {\boldmath$\Downarrow$\hfil}%
                    \typeout{marginpar: $\downarrow$}\ignorespaces}
\def\Mla{\marginpar[\boldmath\hfil\textcolor{red}{$\Rightarrow$}]%
                   {\boldmath\textcolor{red}{$\Leftarrow $}\hfil}%
                    \typeout{marginpar: $\leftrightarrow$}\ignorespaces}
\overfullrule 5pt
\oddsidemargin 15mm
\marginparwidth 29mm
}

\begin{document}

\title{\hfill ~\\[-30mm]
\phantom{h} \hfill\mbox{\small FR-PHENO-2022-01}
\\[1cm]
\vspace{13mm}   \textbf{Quantum integration \\
of elementary particle processes}}

\date{}
\author{
Gabriele Agliardi$^{1,2\,}$\footnote{E-mail:  \texttt{gabrielefrancesco.agliardi@polimi.it}},
Michele Grossi$^{3\,}$\footnote{E-mail:  \texttt{michele.grossi@cern.ch}},
Mathieu Pellen$^{4\,}$\footnote{E-mail:  \texttt{mathieu.pellen@physik.uni-freiburg.de}},
Enrico Prati$^{5,6\,}$\footnote{E-mail:  \texttt{enrico.prati@unimi.it}}
\\[9mm]
{\small\it $^1$ Dipartimento di Fisica, Politecnico di Milano,} \\ %
{\small\it Piazza Leonardo da Vinci 32, I--20133 Milano, Italy}\\
{\small\it $^2$ IBM Italia S.p.A.,} \\ %
{\small\it Via Circonvallazione Idroscalo, I--20090 Segrate (MI), Italy}\\[3mm]
{\small\it $^3$  CERN, 1 Esplanade des Particules, Geneva CH--1211, Switzerland}\\[3mm]
{\small\it $^4$ Albert-Ludwigs-Universit\"at Freiburg, Physikalisches Institut,} \\ %
{\small\it Hermann-Herder-Stra\ss e 3, D--79104 Freiburg, Germany}\\[3mm]
{\small\it $^5$ Istituto di Fotonica e Nanotecnologie, Consiglio Nazionale delle Ricerche,} \\ %
{\small\it Piazza Leonardo da Vinci 32, I--20133 Milano, Italy}\\
{\small\it $^6$ Dipartimento di Fisica Aldo Pontremoli, Università degli Studi di Milano,} \\ %
{\small\it Via Celoria 16, Milan, I--20133, Italy }\\[3mm]
        }
\maketitle

\begin{abstract}
\noindent

We apply quantum integration to elementary particle-physics processes.
In particular, we look at scattering processes such as $\Pe^+\Pe^- \to q \bar q$ and $\Pe^+\Pe^- \to q \bar q' \PW$.
The corresponding probability distributions can be first appropriately loaded on a quantum computer using either quantum Generative Adversarial Networks or an exact method.
The distributions are then integrated using the method of Quantum Amplitude Estimation which shows a quadratic speed-up with respect to classical techniques.
In simulations of noiseless quantum computers, we obtain per-cent accurate results for one- and two-dimensional integration with up to six qubits.
This work paves the way towards taking advantage of quantum algorithms for the integration of high-energy processes.

\end{abstract}
\thispagestyle{empty}
\vfill
\newpage
\setcounter{page}{1}

\tableofcontents
\newpage

\section{Introduction}

In this work, the quantum versions of Monte Carlo algorithms are applied to the problem of integrating elementary-particle cross sections.

In particle physics, integration methods and Monte Carlo programs play a very special role as they are the central link between theory and experiment.
On the one hand, they allow the encoding of theoretical predictions including higher-order or beyond-the-Standard-Model effects. On the other hand, by generating \emph{theoretical} events according to the underlying distribution, they allow a one-to-one correspondence with \emph{experimental} events.
Hence theoretical predictions can be directly compared to experimental measurements in order to get insight into elementary interactions.

For collider experiments such as the Large Hadron Collider (LHC), Monte Carlo simulations are crucial as they simulate all the scattering processes generated in the experiment.
It means that considerable computing resources are needed and they are expected to increase further \cite{Buckley:2019wov,HSFPhysicsEventGeneratorWG:2020gxw}.
For some analysis, the limited Monte Carlo statistics is even becoming a significant source of uncertainty \cite{ATLAS:2019thr,CMS:2019qfk}.
This calls for a continuous improvement of the performance of such Monte Carlo generators.

It appears therefore particularly timely to apply quantum versions of Monte Carlo algorithms to this problem, given the promising advancements in the industry of quantum devices.
The core algorithm of interest for us is Quantum Amplitude Estimation (QAE) \cite{Brassard:2000,Grinko:2019,Suzuki:2019,Nakaji:2020}, that was proven to provide a speedup for the integration of probability distributions, by scaling as $\mathcal{O}\left(1/M\right)$, where $M$ is the number of (quantum) samples, as opposed to classical integrators scaling as $\mathcal{O}(1/\sqrt{M})$ with $M$ (classical) samples.
In the context of high-energy physics, this would translate into a gain in simulation performance, similarly to what was already assessed in other application fields, and specifically in finance  \cite{Rebentrost2018QuantumCF,Zoufal:2019,Stamatopoulos:2020xez,Stamatopoulos:2021eyd}.
%

To apply these techniques, classical data must be loaded into a quantum computer, which is a nontrivial task in terms of computational cost. More precisely, the quantum states that correspond to the data have to be prepared.
In order to encode the data, several algorithms and techniques are used in the literature \cite{grover2002creating,adedoyin2018quantum,woerner2019quantum,gacon2020quantum,holmes2020efficient,garcia2021quantum,Zoufal:2019}.

While the present work is the first application of QAE algorithms to integration, there have been numerous applications of quantum computing to other aspects of collider physics.
Such applications have been mainly experiment-oriented: pixel images \cite{Chang:2021ufg}, event topologies \cite{Kim:2021wrr}, event classification \cite{Bargassa:2021jmk}, Higgs analysis \cite{Belis:2021zqi}, background suppression \cite{Heredge:2021vww}, measurement unfolding \cite{Cormier:2019kcq} or jet clustering \cite{Wei:2019rqy}.
Applications to parton-distribution functions (PDF) have also been carried out by several groups \cite{Perez-Salinas:2020nem,Li:2021kcs} in a quantum context.
In addition, several investigations of quantum parton shower as well as matrix elements evaluation \cite{Bepari:2020xqi,Ramirez-Uribe:2021ubp} have been carried out \cite{Bauer:2019qxa,Bepari:2020xqi,Williams:2021lvr}.
Finally, in Ref.~\cite{Bravo-Prieto:2021ehz} quantum Generative Adversarial Networks (qGAN) \cite{Zoufal:2019} techniques have been used for the purpose of data augmentation.

Here, we focus on two representative cases ($\Pe^+\Pe^- \to q \bar q$ and $\Pe^+\Pe^- \to q \bar q' \PW$ processes) to illustrate the application of QAE.
A particularly important point is that the functions to be integrated are significantly more complicated to than usual Gaussian or log-normal distributions.
These are typically made of trigonometric functions, polynomials, and logarithms (at least for what concerns the lowest order in perturbative theory).
We therefore explore two methods to prepare the quantum states according to the underlying distribution, namely the qGAN \cite{Zoufal:2019} and an exact loading \cite{shende2006synthesis}, respectively.
Of particular interest for this application, is the ability to provide correct results when restricting the domain of integration.
Finally, we carry out one- and two-dimensional integration of cross sections.
For the latter case, we devise a method that is extendable to $n$ dimensions while still allowing the arbitrary reduction of the integration domain.
In general, the integrations are accurate at the per-cent level with up to six qubits.

The article is organised as follows:
in the first part, the method and tools used for this work are presented.
The second part deals with two methods to load the probability distributions.
The third one focuses on integrating such probability distributions with quantum amplitude estimate methods.
Finally, the last sections contains a brief summary as well as some concluding remarks.

\section{Method and tools}

In this section, we first recall some general considerations about Monte Carlo integration and explain how we translate it to our problem.
Second, we briefly describe the processes under investigation.
Finally, the tools used in the next Sections are presented as well as the numerical input of the cross sections.

\subsection*{General considerations}

To start, let us recall some basics of particle physics.
A Monte Carlo integration aims at estimating the cross section of scattering processes which can be written schematically as
\begin{align}
\label{eq:xsection}
 \sigma = \frac{1}{F} \int \rd \Phi \left|\mathcal{M}\right|^2 \Theta(\Phi-\Phi_c),
\end{align}
where $F$ is the flux factor, $\rd \Phi$ the phase-space factor [possibly including parton-distribution function (PDF)],
and $\left|\mathcal{M}\right|^2$ the matrix element squared which encodes the quantum mechanical process.
In addition, the phase space (also called \emph{integration domain} below) can be restricted by the use of so called phase-space cuts which is represented in Eq.~\eqref{eq:xsection} by $\Theta(\Phi-\Phi_c)$ which we refer to as the \emph{domain function} in the following.

In particular, the integration is performed over variables that allow to describe the full phase space.
While these are not physically observable, they allow the full reconstruction of the event kinematic.
In the following, the results are only expressed in terms of these variables that serve as proxies for physical ones.
In particular, the domain restriction (or event selection) is only applied to the variables of integration.
To obtain a \emph{physical} restriction of the domain of integration, a simple mapping can be performed.
In more general terms, any integral $I$ can be cast into the following form
\begin{equation}
 I = \int \rd x f(x) g(x) .
\end{equation}
The function $f$ describes the probability distribution, while the function $g$ is the integrand function. In the QAE, $f$ is computed classically, while $g$ is represented by means of a quantum operator.
For example, in Ref.~\cite{Zoufal:2019}, the $g$ function is a linear function which represents the payoff.
In our case, referring to Eq.~\eqref{eq:xsection}, we take $f=\left|\mathcal{M}\right|^2$.
We then take  $g=\Theta(\Phi-\Phi_c)$, so that $g$ is a generalised Heaviside function which only takes the value 1 or 0; such a function is sometimes called the \textit{indicator function} over the integration domain $D$, and denoted by $\chi_D$ or $\mathbf{1}_D$.\footnote{
	In principle one may also take $f = 1$ and $g = \left|\mathcal{M}\right|^2 \Theta(\Phi-\Phi_c)$, thus eliminating a costly classical pre-computation. Nonetheless, the implementation complexity on the quantum side rises, and more importantly, the quantum circuit becomes deeper and wider, meaning that it could not run on currently available quantum hardware. Consequently we focused our proof of principle on a simplified scenario.
}

\begin{figure}
\center
\includegraphics[width=1\linewidth]{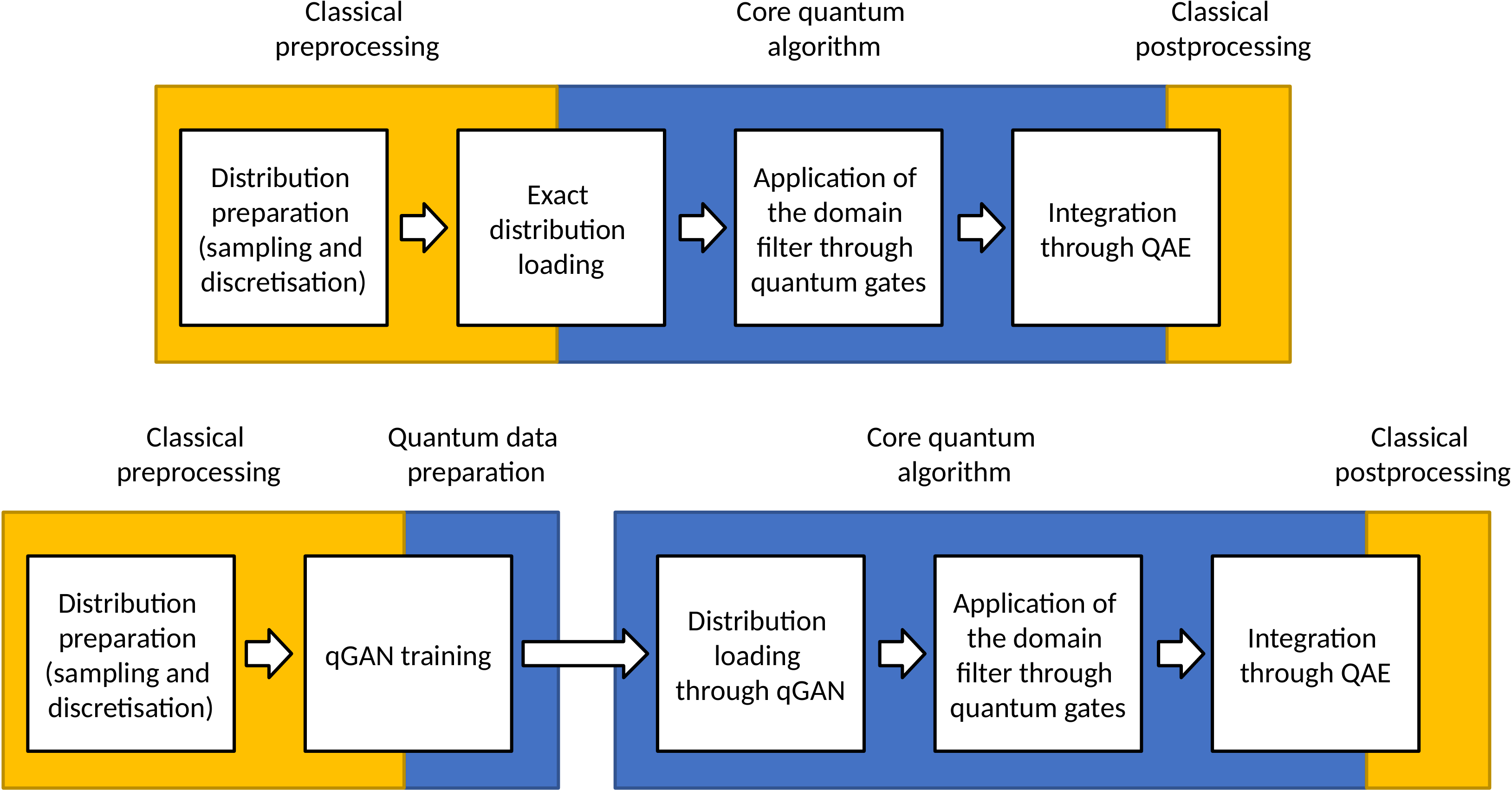}
\caption{Graphical representation of the two approaches followed in this work.
The upper one uses an exact loading method while the lower one is based on the qGAN.}
\label{fig:workflow}
\end{figure}

Implementing this procedure on a quantum computer involves in general two main steps:
the definition of the quantum states and the integration of the probability distribution.
The two approaches that we follow in this work are graphically represented in Fig.~\ref{fig:workflow}.
The first one is based on an exact loading while the second relies on the qGAN to prepare the quantum states.
The details of the implementations are explained in the relevant Sections below.

\subsection*{Particle processes investigated}

In order to test our numerical approach with the quantum simulations, we have focused on two simple though non-trivial scattering processes $\Pe^+\Pe^- \to q \bar q$ and $\Pe^+\Pe^- \to q \bar q' \PW$.
In particular, we have not considered hadronic processes as these would require the use of parton distribution functions.

The first process is $\Pe^+\Pe^- \to q \bar q$.
 In quantum electrodynamics (QED), this process is rather simple.
 By parametrising the phase space with two angles, the cross section reads:
 \begin{equation}
  \sigma \sim \int^{1}_{-1} \int^{2\pi}_0 \rd \cos \theta \rd \phi \left( 1+\cos^2 \theta\right) .
  \label{eq:Xsection22}
 \end{equation}
 This means that computing such a process (up to an overall normalisation factor) simply amounts to integrate the function $1+x^2$ on the integration domain $\left[-1; 1\right]$ while there is no dependence on $\phi$.
 
 
 %
  The second one is $\Pe^+\Pe^- \to q \bar q' \PW$.
  In this case, we have considered the full electroweak Standard Model and not only QED.
  Due to the three particles in the final state, this process has 5 variables of integration.
  These can be chosen as two invariants and three angles and the cross section becomes \cite{byckling1973particle}
  
 \begin{equation}
  \sigma \sim \int^{s}_{M_\PW^2} \int^{s_1^\textrm{Max}}_0 \int^{1}_{-1} \int^{2\pi}_0 \int^{2\pi}_0  \rd \Phi_3 \left| \mathcal{M}_{\Pe^+\Pe^- \to q \bar q' \PW}\right|^2 ,
  \label{eq:Xsection23}
 \end{equation}
 with $s_1^\textrm{Max} = \left(s_2-M_\PW\right)\left(s-s_2\right)/s_2$ and $\rd \Phi_3 = \rd s_2 \rd s_1 \rd \cos \theta_1 \rd \phi_1 \rd \phi_2$.
  As in the previous case, one of the $\phi$ angle is a trivial integration.
  The main characteristics of the process are summarised in Table~\ref{tab:processes}.
  
  \begin{table}[t]
\centering
\begin{tabular}{c|c|c|c}
\multirow{2}{*}{Process number} & \multirow{2}{*}{Description} & Integral & Number of   \\
 & & definition & integration variables \\
\midrule
Process 1 & $\Pe^+\Pe^- \to q \bar q$ & Eq.~\eqref{eq:Xsection22} & $2$   \\
\midrule
Process 2 & $\Pe^+\Pe^- \to q \bar q' \PW$ & Eq.~\eqref{eq:Xsection23} & $5$  \\
\end{tabular}
\caption{Summary of the elementary processes under investigation.}
\label{tab:processes}
\end{table}

\subsection*{Software}

To check our results, we resorted to an in-house Monte Carlo program, that was used for the computation of various high-energy physics processes before \cite{Gavin:2013kga,Gavin:2014yga,Cavasonza:2014xra,Cavasonza:2016qem}.
It is based on the MONACO integration routine which is a modified version of VEGAS \cite{Lepage:1977sw} which is part of the VBFNLO program \cite{Arnold:2008rz,Baglio:2011juf,Baglio:2014uba}.
For the matrix elements, we use either analytical expressions or the matrix-element generator {\sc Recola} \cite{Actis:2012qn,Actis:2016mpe}.

Instead, the results in this article are obtained from the open-source distribution {\sc Qiskit} \cite{Qiskit} which is written in Python. Starting from its libraries, we developed our code to load events, build probability distributions, and calculating integrals.
The specific functions used are described below in the relevant Sections.
With Qiskit, the IBM Quantum Services offer the possibility to run algorithms on simulated quantum computer as well as test some specific configurations on real quantum chips.

\subsection*{Input parameters}

In order to ease reproduction of our results, we provide below the numerical inputs of our simulations.
For the centre-of-mass energy, we have used $\sqrt{s}=500\GeV$.
The electromagnetic coupling is defined with the help of the $G_\mu$ scheme \cite{Denner:2000bj} which leads to
\begin{equation}
  \alpha = \frac{\sqrt{2}}{\pi} G_\mu \MW^2 \left( 1 - \frac{\MW^2}{\MZ^2} \right)  \qquad \text{with}  \qquad   {\GF    = 1.16638\times 10^{-5}\GeV^{-2}}.
\end{equation}
The masses and widths of the massive particles are chosen as
\cite{Tanabashi:2018oca}
\begin{alignat}{2}
\label{eqn:ParticleMassesAndWidths}
                \MZOS &=  91.1876\GeV,      & \quad \quad \quad \GZOS &= 2.4952\GeV,  \nonumber \\
                \MWOS &=  80.379\GeV,       & \GWOS &= 2.085\GeV,  \nonumber \\
\end{alignat}
All other fermions are considered massless.
The pole masses and widths of the heavy gauge bosons are determined from the measured on-shell (OS) values \cite{Bardin:1988xt} via
\begin{equation}
        M_V = \frac{\MVOS}{\sqrt{1+(\GVOS/\MVOS)^2}}\,,\qquad  
\Gamma_V = \frac{\GVOS}{\sqrt{1+(\GVOS/\MVOS)^2}}.
\end{equation}

\section{Definition of probability distributions}

A necessary step that enables the usage and exploitation of a quantum algorithm, is the encoding of data into quantum states, by means of a quantum circuit.
Today, it is not possible to rely on any quantum native techniques like QRAM \cite{qram}. Hence, to solve this potential bottleneck, several approaches were proposed in the literature that allow to encode classical data into quantum states \cite{schuld2021}. 
This is particularly important as the approximation introduced in data loading could affect the quality of the integration.
This procedure corresponds to the first steps depicted in Fig.~\ref{fig:workflow}.

To investigate it, we have classically generated samples (here $10,000$ events) to be loaded into the quantum state.
In particular, we have used two methods: qGAN \cite{Zoufal:2019} and an exact loading \cite{shende2006synthesis}, respectively.
Both approaches will be outlined and compared in the following.
In this Section, we focus on the simple case of $\Pe^+\Pe^- \to q \bar q$ in QED which amounts to integrate $ \int^{+1}_{-1} \rd x \left( 1+x^2\right)$, meaning that the distribution to be loaded is $1+x^2$.

We discuss first the qGAN method for our application. 
A Generative Adversarial Network (GAN) \cite{goodfellow2014generative}, in its classical form, is characterised by the interplay of a generative network and a discriminative network to learn the probability distribution underlying the training data \cite{gui2021review}.
A qGAN has a similar structure, but the generator is a parametrised quantum circuit (PQC) instead of a classical neural network.
This way the generator is trained to load a quantum state, approximating the discretised version of the target distribution.
As a consequence, this algorithm belongs to the general class of quantum variational algorithms, namely hybrid algorithms that rely on a continuous interaction between a quantum computer and a classical computer.
An initial PQC is defined (called \textit{ansatz}) and then using a classical optimiser this circuit is trained iteratively.
The update of the parameters is driven by the evolution of the related loss function. There are no general prescriptions about the structure of the variational circuit, so that challenges remain, including the trainability, accuracy and efficiency of any variational quantum circuits.
For a general overview of variational algorithms, we refer the reader to Ref.~\cite{variatalgo2021}.

\begin{figure}
\center
\includegraphics[width=0.49\linewidth,page=1]{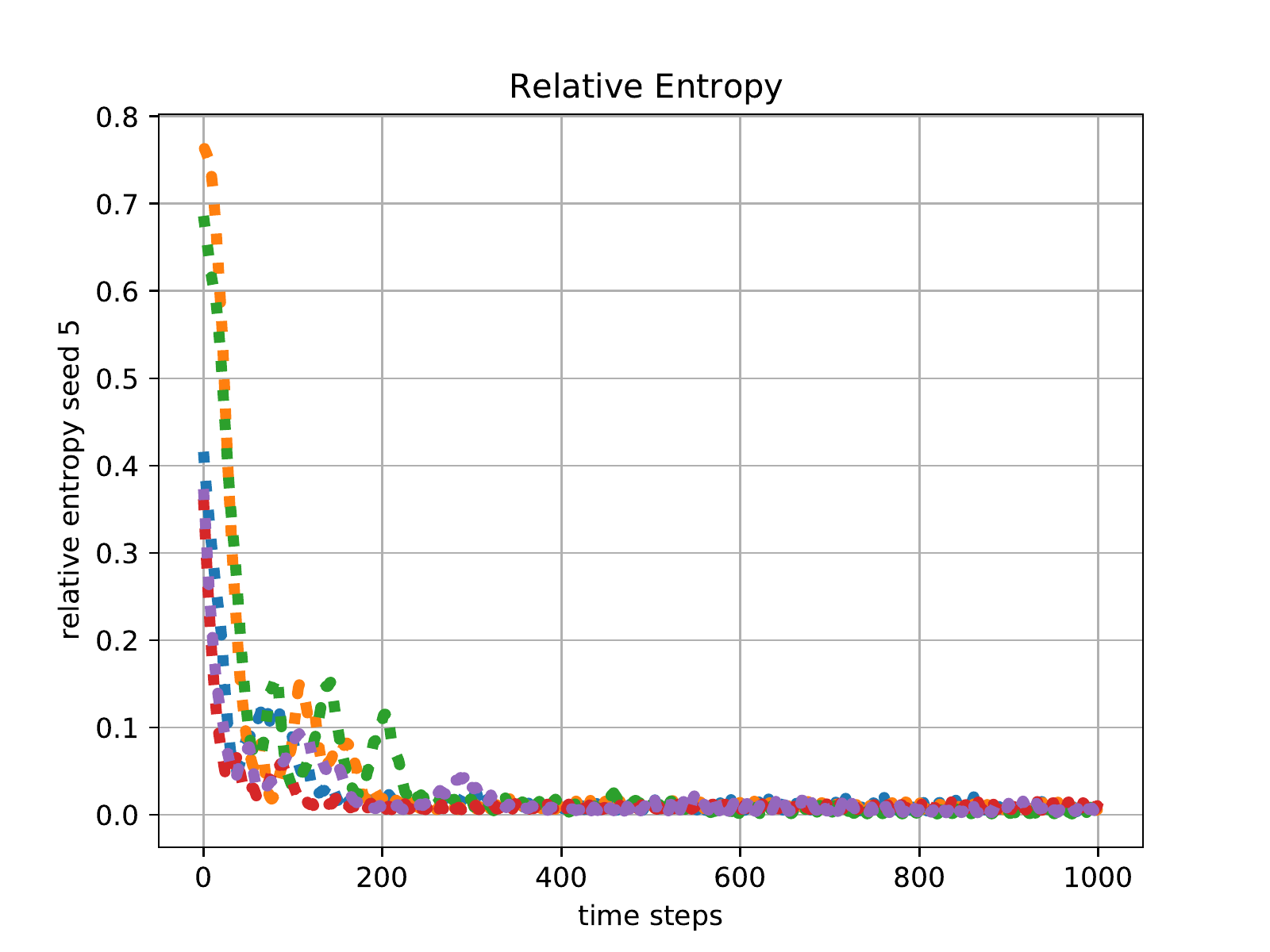}
\includegraphics[width=0.49\linewidth,page=1]{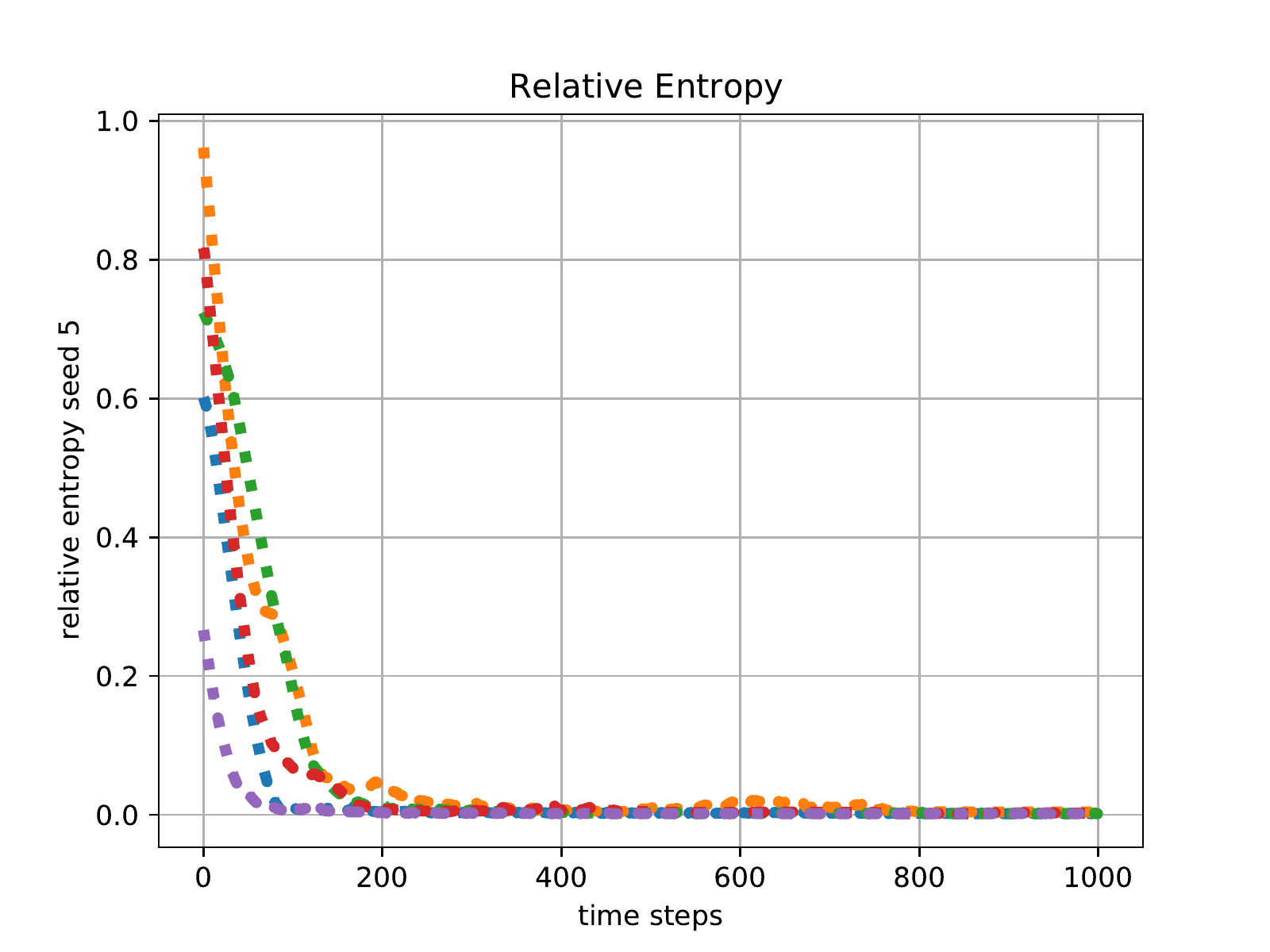}
\includegraphics[width=0.49\linewidth,page=3]{plots/qGAN_1px2_qubits_3_Ep_1000_circular_rep_1_seed1234_default}
\includegraphics[width=0.49\linewidth,page=3]{plots/qGAN_1px2_qubits_3_Ep_1000_circular_rep_1_seed1234_optimised}
\caption{Loading with qGAN of the normalised $1+x^2$ distribution with the default learning rate (left) and an optimised one (right).
In both cases, it is compared to the compared to the theoretical value (thick orange curve) and the entanglement is circular.}
\label{fig:comparison}
\end{figure}

To apply such a method to the integration of the $\Pe^+\Pe^- \to q \bar q$ cross section, we have loaded the normalised distribution $1+x^2$, which we define as $p(x) = (1+x^2)\frac38$ such that $\int^{+1}_{-1} {\rm d} x p(x) = 1$, using the implementation of the qGAN in Qiskit \cite{Zoufal:2019}.\footnote{See for example \url{https://Qiskit.org/documentation/tutorials/finance/10_qgan_option_pricing.html} for the original implementation of Ref.~\cite{Zoufal:2019}.}
The results obtained are presented in Fig.~\ref{fig:comparison} and Table~\ref{tab:comparisonqGAN} where several loaded distributions are compared against the true value of the distribution for two cases.
The first one is the default loading obtained from default qGAN parameters defined in Qiskit, while the second one is an optimised version of the neural network for this particular functional form obtained after several tests of different variational forms and optimiser parameters.

In both cases, five random seeds have been used to estimate the spread of the loading procedure.
From this example, it should be rather clear that an optimisation of the neural network in terms of architecture (rotation gates and entanglement gates) as well as parameters tuning is needed and that a default configuration cannot be used for arbitrary distributions.
Specifically, from our study, the best entanglement is the
\emph{circular} one and the best results are obtained with a learning rate of $5.10^{-4}$ and $1.10^{-3}$ for the generator and
discriminator, respectively.\footnote{The default values of the learning rate for the generator and
discriminator are $1.10^{-3}$ and $1.10^{-5}$, respectively.}
The improvement in the accuracy of the loading can be observed in Fig.~\ref{fig:comparison} and Table~\ref{tab:comparisonqGAN}.
Other strategies for the entanglement layers (\emph{full}, \emph{linear}, or \emph{SCA}) give rather unstable results depending on the seeds used.
Increasing the number of repetitions does not appear to improve the loading accuracy.

In our example, the default qGAN can lead to loading errors up to $40\%$ with an average deviation above $10\%$ per bin.
In the case of an optimised neural network, the average accuracy of the loading per bin is significantly better and lies around $5\%$.
To measure the quality of the whole loaded distribution, one can revert to the root mean squared error defined as
\begin{align}
 \sigma_x = \sqrt{\frac{1}{N} \sum^N_{i=1} (x_i - \mu_i)^2 } ,
\end{align}
where $i$ denotes the bins, $x_i$ the value of the distributions loaded, and $\mu_i$ the true value of the distribution.

Finally, the better behaviour of the tuned qGAN can be also observed in the relative entropy as a function of the time steps in Fig.~\ref{fig:comparison}.
The relative entropy $S$ is defined as
\begin{equation}
 S = \sum_{x=0}^{N-1} P(x) \log \frac{P(x)}{Q(x)} ,
\end{equation}
where $P$ and $Q$ are the output distribution of the quantum generator and the discretised version of the target distribution, respectively.
The tuned network shows a much smoother converge than the default one.

\begin{table}[t]
\centering
\begin{tabular}{c|ccc|c}
\multirow{2}{*}{qGAN loading}  & \multicolumn{3}{c|}{Difference per bin [$\%$]}             & \multirow{2}{*}{$\sigma_x$} \\
 & Min. & Max. & Average \\
\midrule
Default 1 & $+3.46$ &$-25.1$ & $14.6$ & $0.0206$   \\
Default 2 & $+3.90$ &$+19.3$ & $12.0$ & $0.0152$   \\
Default 3 & $+2.36$ &$-21.1$ & $8.51$ & $0.0118$   \\
Default 4 & $+1.48$ &$-40.2$ & $13.7$ & $0.0230$   \\
Default 5 & $+0.224$ &$-31.7$ & $12.0$ & $0.0171$   \\
\midrule
Optimised 1 & $-0.351$ & $-10.0$   & $4.70$ & $7.13\times 10^{-3}$ \\
Optimised 2 & $-0.811$ & $-18.1$   & $7.69$ & $0.0121$ \\
Optimised 3 & $-0.052$ & $-10.1$   & $4.92$ & $7.83\times 10^{-3}$ \\
Optimised 4 & $+0.599$ & $-15.4$   & $5.16$ & $7.64\times 10^{-3}$ \\
Optimised 5 & $-0.995$ & $-12.4$   & $4.65$ & $7.00\times 10^{-3}$ \\
\end{tabular}
\caption{Comparison of qGAN loading of the normalised $1+x^2$ distribution for the default learning rates and an optimised one. The results are given for 5 different seeds.
The minimum, maximum, and average difference per bin with respect to the true value is provided (in per cent). The root mean squared error from the true value is also given.}
\label{tab:comparisonqGAN}
\end{table}

\begin{figure}
\center
\includegraphics[width=0.5\linewidth]{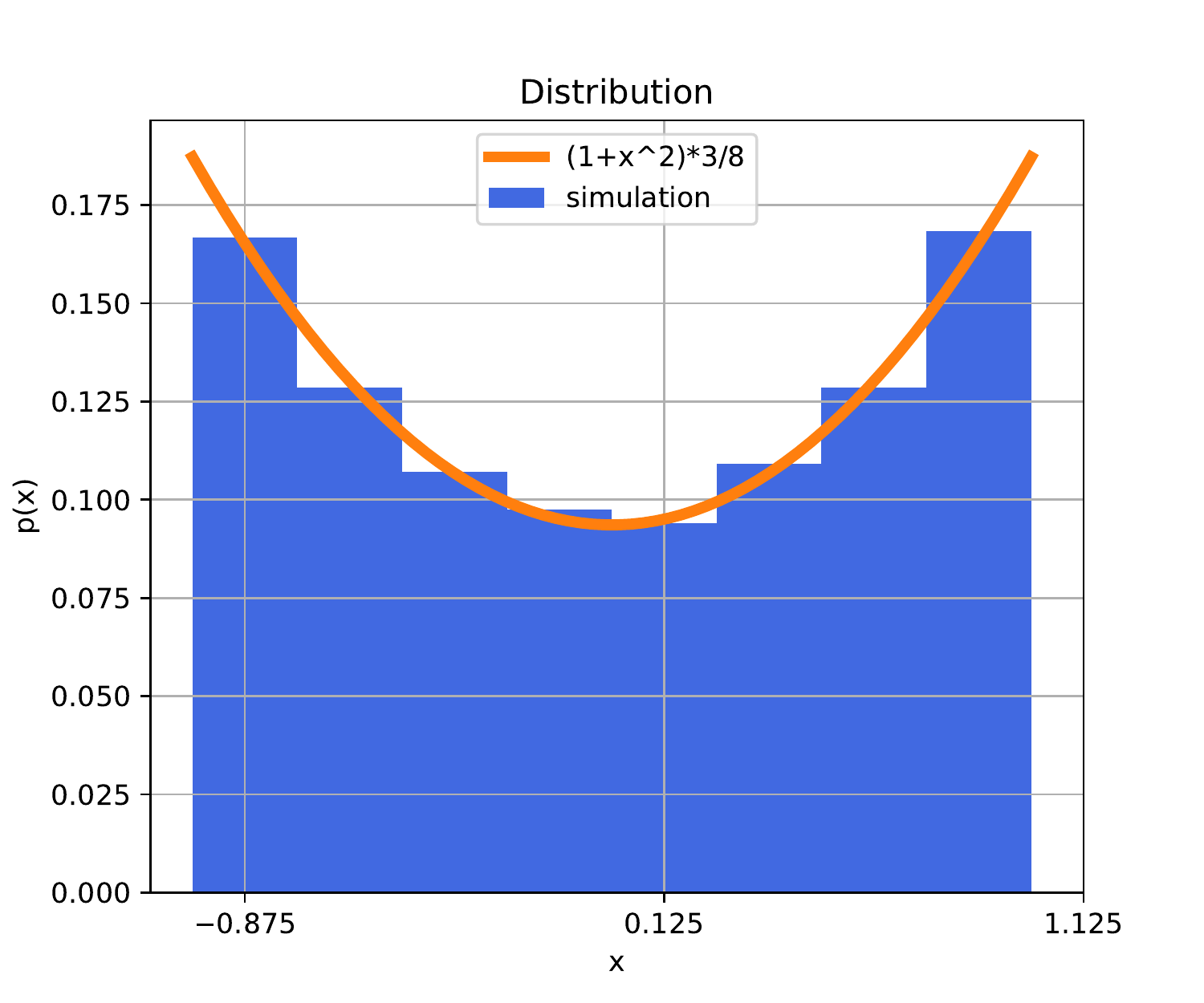}
\caption{Direct loading of the the normalised $1+x^2$ distribution (blue histogram) compared to the theoretical value (orange curve).}
\label{fig:loading}
\end{figure}

We now turn to the \emph{exact loading} which is represented in Fig.~\ref{fig:loading}.
Such a technique is an analytical way to initialise complex amplitude on qubit register.
Being an exact method, the accuracy of the loaded distribution is obviously better than with the qGAN.
This is shown quantitatively in Table~\ref{tab:comparisonLoading} where the differences per bin are shown for the best bin, the worst, and the average.
For the two qGAN cases above, the best seed has been selected.
From Table~\ref{tab:comparisonLoading}, it is rather clear that the exact loading is performing significantly better in this case.
In particular, it shows discrepancies from the truth by no more than $2\%$ and is on average around $1\%$.
This order of magnitude should be kept in mind as the best possible loading accuracy with a sample of $10,000$ events.

\begin{table}[t]
\centering
\begin{tabular}{c|ccc|c}
\multirow{2}{*}{Loading}  & \multicolumn{3}{c|}{Difference per bin [$\%$]} & \multirow{2}{*}{$\sigma_x$} \\
 & Min. & Max. & Average \\
\midrule
    Direct         & $+0.207$ & $-1.88$ & $1.35$ & $1.80\times 10^{-3}$ \\
    qGAN default   & $+2.36$   & $-21.1$ & $8.51$  & $0.0118$   \\
    qGAN optimised & $-0.995$  & $-12.4$ & $4.65$  & $7.00\times 10^{-3}$ \\
\end{tabular}
\caption{Comparison of qGAN loading of the normalised $1+x^2$ distribution for the default learning rates and an optimised one as well as the direct loading.
The qGAN results are the ones of the best seed in table \ref{tab:comparisonqGAN}.
The minimum, maximum, and average difference per bin with respect to the true value is provided (in per cent). The root mean squared error from the true value is also given.}
\label{tab:comparisonLoading}
\end{table}

If instead, one loads the exact distribution which is known analytically (as opposed to a sample generated according to it), there is simply no deviation from the true distribution.
This aspect is particularly important as in principle, a closed form of the distribution is not necessarily available.
It also means that the quality of the exact loading is directly dependent on the statistics of the sample given as input.

In order to represent the target distributions with $N$ bins, one needs to encode a statevector of size $N$, and this translates into a number of qubits $n$ such that $N=2^n$. In other words, the data resolution is a direct consequence of the number of qubits used.
Obviously, the computational complexity is related to such $n$, as well.
As far as the qGAN approach is concerned, and discarding the training process that will be discussed in the next paragraph, the pure loading phase requires $\mathcal{O}(poly(kn))$ gates, where $k$ is the number of layers, which is intrisic in the definition of the ansatz. Assuming that $k$ can be kept under control, qGANs become an efficient data loading technique, and preserve the speedup of the Quantum Amplitude Estimation algorithm for integration \cite{Zoufal:2019}. 
Conversely, for the \emph{exact loading} algorithm, the number of 2-qubit gates scales as $\mathcal{O}(2^n)$ \cite{shende2006synthesis}.

In the argument above, the training cost of a qGAN is neglected. This is motivated by the fact that the same distribution is typically used for multiple simulations, and in this case the training process is performed once, so that the training time can be seen as a constant. Nonetheless, it is worth saying that the scaling of the training cost when the distribution size grows, is an open question, whose complexity lies in the unpredictable number of epochs needed to achieve training convergence, in the desired level of approximation, and in the different behavior of various optimisers. The interaction of such hyperparameters on small-scale problems is discussed in Ref.~\cite{agliardi2022optimal}.

Given the limited amount of qubits in our study, we could not appreciate the benefits of qGANs in terms of scaling, while on the contrary we had to face the learning, possibly the tuning of the network, and the verification of the result.
Moreover, the current absence of analytical estimates for approximations induced by qGANs, limits their applicability for computing arbitrary processes in a quantum Monte Carlo program, especially when probability distributions are known from first principle or vast amounts of classical data representing them are available.

\section{Integration of probability distributions}

In this section, we exclusively discuss the integration of probability distributions.
While we could also use the qGAN loading, we use exclusively the exact loading method in this Section in order to isolate the integration step from the loading one.
In particular, we look at the integration of one- and two-dimensional distributions.
This procedure corresponds to the core of the work flow depicted in Fig.~\ref{fig:workflow}.

Once the target distribution has been loaded into a quantum channel, the integration is performed through QAE. 
Assuming efficient data loading, the algorithm achieves a quadratic improvement, compared with classical Monte Carlo simulation. QAE is a very interesting and studied quantum algorithm due to its potential application in different fields such as quantum chemistry, machine learning, finance and high energy physics.
QAE is a fundamental routine in quantum computing which generalises the idea behind the Grover's search algorithm,
 and gives rise to a family of quantum algorithms. The basic idea is that given an operator $\mathcal{A}$ that acts as \\
 \begin{equation}
 \mathcal{A}|0\rangle = \sqrt{1 - a}|\Psi_0\rangle + \sqrt{a}|\Psi_1\rangle
 \end{equation}
 where $a \in [0,1]$ and $| \Psi_0\rangle$ and $| ex\Psi_1\rangle$ are two normalised states.
Quantum Amplitude Estimation (QAE) is the task of finding an estimate for the amplitude $a$ of the state $|\Psi_1\rangle$:
\begin{equation*}
a = |\langle\Psi_1 | \Psi_1\rangle|^2.
\end{equation*}
This can be achieved by the definition of a Grover's like operator of the form \cite{Brassard:2000}:
\begin{equation}\label{eq:qoperator}
\mathcal{Q} = \mathcal{A}\mathcal{S}_0\mathcal{A}^\dagger\mathcal{S}_{\Psi_1} ,
\end{equation}
where $\mathcal{S}_0$ and $\mathcal{S}_{\Psi_1}$ are reflections about the $|0\rangle$ and $|\Psi_1\rangle$ states, respectively, and phase estimation.
 This formulation represents the canonical version of QAE which is a combination of Quantum Phase Estimation (QPE) and Grover’s Algorithm \cite{nielsen_chuang_2010}.
 On one hand, QPE is theoretically able to achieve exponential speedup, like in the famous Shor's Algorithm for factoring \cite{shor1999polynomial},
 on the other hand its practical implementation in terms of qubits and circuit depth represents an interesting challenge in current technological scenario.
 Removing the dependence on QPE for a QAE-like routine in a simplified version such that it uses only Grover iterations has been largely studied in the literature.
 
 Indeed, there exist different implementations, with respect to the original QAE implementation by Brassard et al. \cite{Brassard:2000},
 such as the Iterative Amplitude Estimation (IAE) version which does not rely on Quantum Phase Estimation (QPE) as defined in Eq.\eqref{eq:qoperator}.
 This is the adopted version for this work which can achieve a provable quadratic speedup over classical Monte Carlo simulation, with a desired asymptotic
 behaviour in its iterative queries to the quantum computer, reducing the required number of qubits and gates \cite{Grinko:2019}.
 Additional implementations are the Maximum Likelihood Amplitude Estimation \cite{Suzuki:2019,Nakaji:2020} which limit resorting to expensive
  controlled operations.

\subsection*{One-dimensional distribution}

As mentioned in the previous section, the direct loading adds no approximation to the probabilities given as input. If such probabilities are obtained through sampling, though, they are in turn approximated.
This means eventually that the result of the integration will strongly depend on the quality of the input.
To illustrate this, we have use the QAE with samples of different sizes: 1000 events (low statistics), 100,000 events (high statistics), 1M events (very high statistics).
In particular, we have made used of the Qiskit functions {\sc LinearAmplitudeFunction} \cite{woerner2019quantum,gacon2020quantum}, {\sc EstimationProblem}, and {\sc IterativeAmplitudeEstimation}.
The latter implement and improved version of the original QAE method \cite{Grinko:2019}.
The results for the different samples and the loading of the exact distribution are compared to the analytical result in Tables~\ref{tab:integrationSym} and \ref{tab:integrationAsym}.
In these Tables and the following ones, $\delta [\%] = \frac{\sigma-\sigma_{\rm truth}}{\sigma_{\rm truth}}$ in per cent, where $\sigma_{\rm truth}$ denotes the true analytical integration.

\begin{table}[t]
\centering
\begin{tabular}{c|cc|cc|cc|cc}
\multirow{2}{*}{Domain}  & \multicolumn{2}{c|}{low stat.} & \multicolumn{2}{c|}{high stat.} & \multicolumn{2}{c|}{very high stat.} & \multicolumn{2}{c}{exact} \\
 & $\sigma$ & $\delta [\%]$ & $\sigma$ & $\delta [\%]$ & $\sigma$ & $\delta [\%]$ & $\sigma$ & $\delta [\%]$ \\
\midrule
$[-0.75; 0.75]$ & $0.664$ & $0.592$ & $0.664$ & $0.622$ & $0.668$ & $0.0280$ & $0.668$ & $-2.01\times 10^{-3}$ \\
$[-0.5; 0.5]$ & $0.403$ & $0.794$ & $0.402$ & $1.16$ & $0.406$ & $0.122$ & $0.406$ & $-6.01\times 10^{-3}$ \\
$[-0.25; 0.25]$ & $0.196$ & $-2.42$ & $0.189$ & $1.01$ & $0.192$ & $-0.166$ & $0.191$ & $-0.0175$
 \end{tabular}
\caption{Symmetric integration of the normalised $1+x^2$ probability distribution based on samples with different statistics (low, high, and very high) or the exact probability distribution.
The results are compared to the analytical result in per cent. The results are obtained for three qubits.
The low, high, and very high statistics refer to $10,000$, $100,000$, and $1$ million events, respectively.}
\label{tab:integrationSym}
\end{table}

\begin{table}[t]
\centering
\begin{tabular}{c|cc|cc|cc|cc}
\multirow{2}{*}{Domain}  & \multicolumn{2}{c|}{low stat.} & \multicolumn{2}{c|}{high stat.} & \multicolumn{2}{c|}{very high stat.} & \multicolumn{2}{c}{exact} \\
 & $\sigma$ & $\delta [\%]$ & $\sigma$ & $\delta [\%]$ & $\sigma$ & $\delta [\%]$ & $\sigma$ & $\delta [\%]$ \\
\midrule
$[-0.75; 0]$ & $0.345$ & $-3.31$ & $0.332$ & $0.706$ & $0.334$ & $0.0331$ & $0.334$ & $-8.31\times 10^{-3}$ \\
$[-0.5; 0]$ & $0.215$ & $-5.86$ & $0.201$ & $1.15$ & $0.203$ & $0.0986$ & $0.203$ & $-0.0161$ \\
$[-0.25; 0]$ & $0.112$ & $-17.1$ & $0.0939$ & $1.87$ & $0.0960$ & $-0.284$ & $0.0957$ & $-0.0389$
 \end{tabular}
\caption{Asymmetric integration of the normalised $1+x^2$ probability distribution based on samples with different statistics (low, high, and very high) or the exact probability distribution.
The results are compared to the analytical result in per cent. The results are obtained for three qubits.
The low, high, and very high statistics refer to $10,000$, $100,000$, and $1$ million events, respectively.}
\label{tab:integrationAsym}
\end{table}

It is particularly visible that the quality of the integration is dependent on the statistics used.
For 1 million events, the result of the integration is accurate at around the per-mille level.
The loading of the exact distribution, on the other hand, is systematically below half a per mille accuracy.
In addition, it is worth emphasising that the relevant statistics for the integration precision is not the one of the full sample but of the sample in the integrated region.
This is particularly clear in Tables~\ref{tab:integrationSym} and \ref{tab:integrationAsym} where the smaller integration domain have a lower accuracy.
This holds true also for the the loading of the exact distribution.
It is worth noticing that the relative differences with respect to the true values in Tables~\ref{tab:integrationSym} and \ref{tab:integrationAsym} do not necessarily display a scaling behaviour according to the statistics.
This is due to the fact that the samples are subject to statistical fluctuation and their central value (as opposed to the error) follow a scaling behaviour only on average and not for every single point.
These numbers are particularly useful as they provide an estimate of the error which originates from not knowing the original distribution analytically (as in the 2D case below). In the present case, this error is about few per cent.

\begin{table}[t]
\centering
\begin{tabular}{c|cc|cc|cc|cc}
\multirow{3}{*}{Qubits number}  & \multicolumn{4}{c|}{$[-0.7; 0.7]$} & \multicolumn{4}{c}{$[-0.625; 0.625]$} \\
 & \multicolumn{2}{c|}{high stat.} & \multicolumn{2}{c|}{exact} & \multicolumn{2}{c|}{high stat.} & \multicolumn{2}{c}{exact} \\
 & $\sigma$ & $\delta [\%]$ & $\sigma$ & $\delta [\%]$ & $\sigma$ & $\delta [\%]$ & $\sigma$ & $\delta [\%]$ \\
\midrule
$3$ & $0.402$ & $-34.3$ & $0.406$ & $-33.5$ & $0.402$ & $-24.2$ & $0.406$ & $-23.3$ \\
$4$ & $0.525$ & $-14.1$ & $0.530$ & $-13.2$ & $0.525$ & $-0.933$ & $0.530$ & $3.67\times 10^{-3}$ \\
$5$ & $0.592$ & $-3.05$ & $0.597$ & $-2.27$ & $0.525$ & $-0.933$ & $0.530$ & $3.67\times 10^{-3}$ \\
$6$ & $0.592$ & $-3.05$ & $0.597$ & $-2.27$ & $0.525$ & $-0.933$ & $0.530$ & $3.67\times 10^{-3}$
 \end{tabular}
\caption{Symmetric integration of the normalised $1+x^2$ probability distribution based on a 1 million-events samples as well as the exact probability distribution.
The results are compared to the analytical result in per cent as a function of the number of qubits.
The high statistics refer to $100,000$ events.}
\label{tab:integrationSymQ}
\end{table}

\begin{table}[t]
\centering
\begin{tabular}{c|cc|cc|cc|cc}
\multirow{3}{*}{Qubits number}  & \multicolumn{4}{c|}{$[-0.7; 0.6]$} & \multicolumn{4}{c}{$[-0.625; 0.375]$} \\
 & \multicolumn{2}{c|}{high stat.} & \multicolumn{2}{c|}{exact} & \multicolumn{2}{c|}{high stat.} & \multicolumn{2}{c}{exact} \\
 & $\sigma$ & $\delta [\%]$ & $\sigma$ & $\delta [\%]$ & $\sigma$ & $\delta [\%]$ & $\sigma$ & $\delta [\%]$ \\
\midrule
$3$ & $0.402$ & $-28.0$ & $0.406$ & $-27.1$ & $0.296$ & $-28.1$ & $0.299$ & $-27.5$ \\
$4$ & $0.463$ & $-17.0$ & $0.468$ & $-16.0$ & $0.408$ & $-1.07$ & $0.412$ & $5.96
\times 10^{-3}$ \\
$5$ & $0.527$ & $-5.46$ & $0.532$ & $-4.62$ & $0.408$ & $-1.07$ & $0.412$ & $5.96
\times 10^{-3}$ \\
$6$ & $0.542$ & $-2.76$ & $0.547$ & $-1.81$ & $0.408$ & $-1.07$ & $0.412$ & $5.96
\times 10^{-3}$
 \end{tabular}
\caption{Asymmetric integration of the normalised $1+x^2$ probability distribution based on a 1 million-events samples as well as the exact probability distribution.
The results are compared to the analytical result in per cent as a function of the number of qubits.
The high statistics refer to $100,000$ events.}
\label{tab:integrationAsymQ}
\end{table}

While in Tables~\ref{tab:integrationSym} and \ref{tab:integrationAsym}, the limits of integration corresponds to the eight bins ($n=3$ qubits give $2^n$ bins) on the domain $\left[-1; 1\right]$, in Tables~\ref{tab:integrationSymQ} and \ref{tab:integrationAsymQ} the same exercise is performed with this time integration domains that do not fit the binning of the piecewise definition of the function.

In the present case, only the results of the integration of the high-statistics sample as well as the exact result are provided as a function of the number of qubits.
In general, one observes that the results are significantly worse than in the previous case.
This is simply due to the ill-defined value of the distribution between two bins.
By increasing the number of qubits, one observes an improvement of the results until the bin edges correspond to the integration boundaries.
Once the bin edges fit the integration boundaries, increasing the number of qubits does not lead to any improvement as the distribution is already best defined within the integration boundaries.
This implies that when taking the limit of large numbers of qubits, these artifacts disappear.

\subsection*{Two-dimensional distribution}

\begin{figure}
\center
\includegraphics[width=0.52\linewidth]{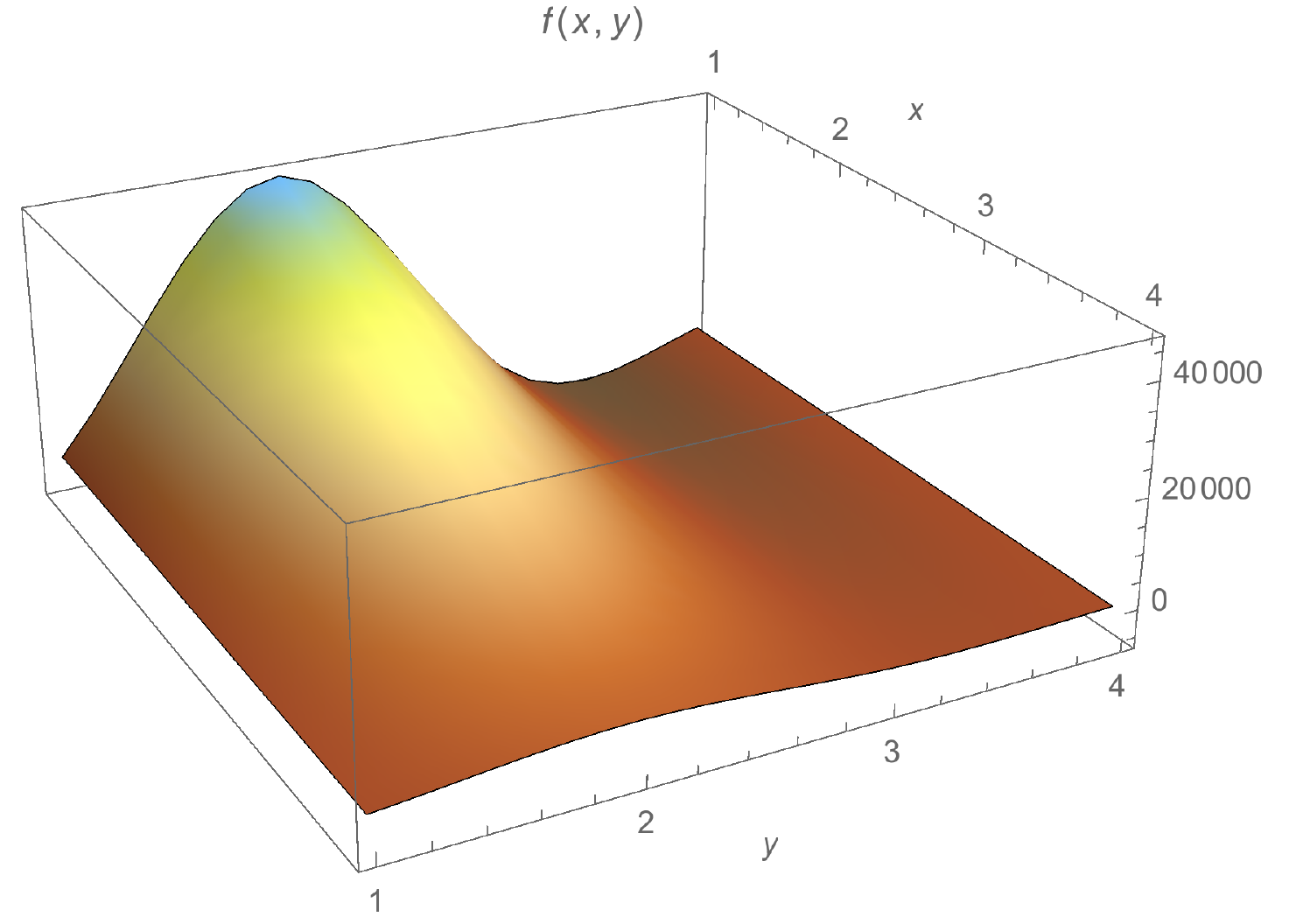}
\includegraphics[width=0.47\linewidth]{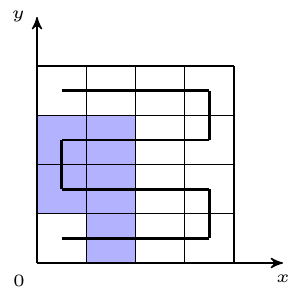}
\caption{Two-dimensional numerical representation of the integrand in Eq.~\eqref{eq:2D} with $x=s_2$ and $y=s_1$ (left).
Graphical representation of the mapping of the two-dimensional function to a one dimensional function (right). The blue shaded area represents a restriction of the domain of integration.}
\label{fig:Snake}
\end{figure}

We now turn to the integration of a two-dimensional function for the case of $\Pe^+\Pe^- \to q \bar q' \PW$.
As it can be seen from Eq.~\eqref{eq:Xsection23}, the 3-particles phase space requires the integration over 5 variables.
To simplify the problem while keeping it non-trivial, we integrate over the two invariants $s_1$ and $s_2$.
To that end, we take a slice in Eq.~\eqref{eq:Xsection23} by setting $\cos \theta_1=0$, $\phi_1 = \pi/2$, and $\phi_2 = \pi/2$.
The cross section then becomes 

 \begin{equation}
  \sigma \sim \int^{s}_{M_\PW^2} \int^{s_1^\textrm{Max}}_0 \rd \tilde\Phi_3 \left| \mathcal{M'}\right|^2,
  \label{eq:2D}
 \end{equation}
 with $\rd \tilde\Phi_3 = \rd s_2 \rd s_1$ and $\mathcal{M'} = \mathcal{M}_{\Pe^+\Pe^- \to q \bar q' \PW}\left(\cos \theta_1=0, \; \phi_1=\pi/2, \; \phi_2=\pi/2 \right)$.
 The integration of the cross section therefore amounts to integrate over the variables $s_2$ and $s_1$.
 The integrand is graphically represented on the left-hand side of Fig.~\ref{fig:Snake} as a function of $x=s_2$ and $y=s_1$.
 Again, we would like to stress that, as in the one-dimensional case, the type of functions to be integrated are rather different and more complicated than those that have been tested so far such as Gaussian or log-normal distributions.

 To encode the multidimensional distribution of Eq.~\eqref{eq:2D} on qubits register, we revert to the same method as in the one-dimensional case by defining it piecewise.
 To that end, we introduce a mapping from the two dimensional function to one-dimensional.
 The way the mapping is performed is represented in the right-hand side of Fig.~\ref{fig:Snake}.
 We opt for this solution instead of simply scanning from left to right and top to bottom, in order to ensure that any physically motivated integration domain restrictions can be mapped to the one-dimensional function in a continuous way, hence minimising the error due to interpolations that are not evident nor considered in previous one-dimensional case.
 We note that this method is fully general and can be extended to $n$-dimensional integral.
 It has also the advantage to be fully flexible and allow for the arbitrary phase-space cut in the integrand.
 For example in Fig.~\ref{fig:Snake}, the blue shaded area represents the restriction of the integration domain.
 In this case, assuming that the first bin of the 1D function [$\tilde f \mapsto \tilde f(X)$] is in the top left corner and that it is mapped to $\mathcal{S}_X =\left[0; 16 \right]$, break points will be introduced at $X=6$, $X=10$, $X=14$, and $X=15$.
 
 For the present experiment, we have produced $100,000$ events according to the two-dimensional distribution over the full integration domain.
 We reduced the latter by setting the maximum value of $s_2$ to $20,000$ in order to avoid populating bins with very few events.
 This leads to a total of $97,581$ events.
  
\begin{table}[t]
\centering
\begin{tabular}{c|c|cc|cc}
Qubits & \multirow{2}{*}{Grid dim.\ } & \multicolumn{2}{c|}{$\mathcal{S}_1$} & \multicolumn{2}{c}{$\mathcal{S}_2$}  \\
number & & $\sigma$ & $\delta [\%]$ & $\sigma$ & $\delta [\%]$  \\
\midrule
$4$ & $4\times4$ & $0.55$ & $0$ & $0.70$ & $-4.1$  \\
$5$ & $5\times5$ & $0.52$ & $-4.92$ & $0.53$ & $-26.6$  \\
$6$ & $6\times6$ & $0.47$ & $-14.1$ & $0.79$ & $9$  \\
$6$ & $7\times7$ & $0.62$ & $-14.4$ & $0.70$ & $-3.0$  \\
$6$ & $8\times8$ & $0.55$ & $0$ & $0.78$ & $7.6$
 \end{tabular}
\caption{Two-dimensional integration with two different integration domains:
one which matches bin edges ($\mathcal{S}_1$) and the other does not ($\mathcal{S}_2$).
The numerical integrations are compared to the value obtained from the classical sample for different grid dimensions corresponding to different number of qubits.
.}
\label{tab:2Dintegration}
\end{table}

The results of our experiments are given in Table~\ref{tab:2Dintegration}.
In this case, we consider two cases of integrand reduction or cuts:
$\mathcal{S}_1$ which implement the cut $\left[62,500; 187,500 \right] \times \left[5,000; 10,000 \right]$ and $\mathcal{S}_2$ which corresponds to $\left[80,000; 150,000 \right] \times \left[5,000; 10,000 \right]$.
The main difference between these two sub-domains is that the boundaries of $\mathcal{S}_1$ fits the edges of the bins of a $4\times4$ grid while the ones of $\mathcal{S}_2$ do not for the first variable.

This explains why in Table~\ref{tab:2Dintegration} for the $4\times4$ grid,
the result of the integration is perfectly reproducing the truth ($\sigma = 0.545833717629457$) which is here the classical sample.
While in the one-dimensional case, each increase in the number of qubits translates into the halving of the bins, it is not the case here.
Indeed, going from 4 qubits to 5, only allows to extend the grid from $4\times4$ to $5\times5$.
It explains why for $\mathcal{S}_1$, while increasing the grid and making the binning finner, the accuracy of the integration does not improve.
It only becomes perfect again when the binning is again perfectly fitting the boundaries of the integration domain.

This is further exemplified with the case of $\mathcal{S}_2$ where the improvement is not uniform when increasing the grid dimension.
For this case, the true value of the cross section is $\sigma = 0.7244852993923$.
This is due, on the one hand, to the fact that the edges of the second dimension are only matched for the cases of the $4\times4$ and $8\times8$ grids.
On the other hand, as seen in the one dimensional case, when the domain of integration does not match the piecewise definition of the function, the result of the integration is uncontrolled.
In the present case, the interpolation is such that doubling the number of bins in each dimension does not necessarily increase the precision of the integration (grid $4\times4$ vs.\ grid $8\times8$).

This implies that only a large number of qubits (implying finer bins) can allow a reliable estimate of the integral.
In particular, for the present application which the computation of cross sections in collider experiments, the usual standard for Monte Carlo error is to reach a per mille accuracy.
With current technology, this goal could be challenging.
Nonetheless, we believe that with the advent of machines with $1000$ qubits or more\footnote{See for example, IBM recent \emph{Roadmap to Scaling Quantum Technology} announcing aiming at 1000-plus qubits by 2023.}, this is perfectly conceivable. Not only a greater number of qubits is needed but also a greater quantum volume \cite{moll2018quantum} that could allow to run QAE on a quantum computer, where further improvements are required, \emph{e.g.}, longer coherence times and higher gate fidelity.

We note in passing that with our method,
we could in principle also sample events according to the underlying distribution as done in Ref.~\cite{Bravo-Prieto:2021ehz}.
We defer the study of this aspect to future work.
In particular, while the integration of probability distributions with QAE methods has shown to provide a quantum advantage, it is not clear yet if such an advantage can also be observed for the sampling of events.

\section{Conclusion and outlook}

This work constitutes the first application of Quantum Amplitude Estimate (QAE) algorithm to high-energy physics.
To test its feasibility we have checked two non-trivial elementary processes, namely $\Pe^+\Pe^- \to q \bar q$ and $\Pe^+\Pe^- \to q \bar q' \PW$.

Complex function appearing in elementary scattering processes can successfully be loaded onto qubits consistently with the results of Ref.~\cite{Bravo-Prieto:2021ehz}.
To load the functions we have used two methods, namely:
the quantum Generative Adversarial Networks (qGAN) \cite{Zoufal:2019} and an exact loading \cite{shende2006synthesis}.
For our purposes, we have found that the latter one is more appropriate due to its versatility and reliability for what concerns application with a small number of qubits.
In particular, it does not require any training nor tuning which makes it very easy to use.

In addition, we have successfully used the QAE algorithm for the integration of the two elementary processes in one and two dimensions, respectively.
In particular, we have tested the reliability of the integration when restricting its domain of integration, which would correspond to imposing physical event selection in an experiment.
To integrate multi-dimensional functions, we have devised a general method which can be extended to $n$ dimensions.

Following this purely numerical strategy requires large number of qubits in order to be accurate.
For our application, we have found that QAE provides per-cent accurate results for one- and two-dimensional integration with up to six qubits.
The results support the framework where future physical devices will make quantum computing a viable solution for integrating elementary processes in high-energy physics. An increase in the number of available qubits is critical for the practical application to our domain of study. It should be noted here, though, that other issues emerge in the current era of Noisy Intermediate-Scale Quantum Computers \cite{preskill_quantum_2018}. Indeed, additional challenges originate from the imperfection of present hardware construction, from the limited topological connectivity of qubits, and from the inability to put in place full error correction protocols that would require additional qubits and resources. Practical usage of the algorithms shall therefore be validated and perfected also through the execution on real quantum devices.

This work opens new perspectives for the computation of particle processes with quantum Monte Carlo integration techniques.
Following the same method, more complicated processes (with higher multiplicities and hadronic processes) can be investigated.
%

\section*{Acknowledgements}

The Authors are grateful to Julien Gacon, Ivano Tavernelli, Sofia Vallecorsa, and Christa Zoufal for useful discussions.
We acknowledge use of the IBM Quantum platform for this work.
This project is supported by CERN Quantum Technology Initiative.
The views expressed are those of the authors and do not reflect the official policy or position of IBM company or the IBM Quantum team.
GA is grateful to IBM for supporting his Executive PhD.
GA and EP are thankful for the access to the IBM Quantum Researchers Program.
MP acknowledges support from the German Research Foundation (DFG) through the Research Training Group RTG2044.

\bibliographystyle{utphys.bst}
\bibliography{qchep}

\providecommand{\href}[2]{#2}\begingroup\raggedright\begin{thebibliography}{10}

\bibitem{Buckley:2019wov}
A.~Buckley, {\em {Computational challenges for MC event generation}}.
  \href{http://dx.doi.org/10.1088/1742-6596/1525/1/012023}{J. Phys. Conf. Ser.
  {\bf 1525} (2020) no.~1, 012023}, \href{http://arxiv.org/abs/1908.00167}{{\tt
  arXiv:1908.00167 [hep-ph]}}.

\bibitem{HSFPhysicsEventGeneratorWG:2020gxw}
{\bf HSF Physics Event Generator WG} Collaboration, S.~Amoroso {\em et al.},
  {\em {Challenges in Monte Carlo Event Generator Software for
  High\nobreakdash-Luminosity LHC}}.
  \href{http://dx.doi.org/10.1007/s41781-021-00055-1}{Comput. Softw. Big Sci.
  {\bf 5} (2021) no.~1, 12}, \href{http://arxiv.org/abs/2004.13687}{{\tt
  arXiv:2004.13687 [hep-ph]}}.

\bibitem{ATLAS:2019thr}
{\bf ATLAS} Collaboration, G.~Aad {\em et al.}, {\em {Search for the
  electroweak diboson production in association with a high-mass dijet system
  in semileptonic final states in $pp$ collisions at $\sqrt{s}=13$ TeV with the
  ATLAS detector}}. \href{http://dx.doi.org/10.1103/PhysRevD.100.032007}{Phys.
  Rev. D {\bf 100} (2019) no.~3, 032007},
  \href{http://arxiv.org/abs/1905.07714}{{\tt arXiv:1905.07714 [hep-ex]}}.

\bibitem{CMS:2019qfk}
{\bf CMS} Collaboration, A.~M. Sirunyan {\em et al.}, {\em {Search for
  anomalous electroweak production of vector boson pairs in association with
  two jets in proton-proton collisions at 13 TeV}}.
  \href{http://dx.doi.org/10.1016/j.physletb.2019.134985}{Phys. Lett. B {\bf
  798} (2019)  134985}, \href{http://arxiv.org/abs/1905.07445}{{\tt
  arXiv:1905.07445 [hep-ex]}}.

\bibitem{Brassard:2000}
G.~Brassard, M.~Mosca, and A.~Tapp, {\em {Quantum Amplitude Amplification and
  Estimation}}. \href{http://dx.doi.org/10.1090/conm/305/05215}{Quantum
  Computation and Information {\bf 305} (2002)  },
  \href{http://arxiv.org/abs/quant-ph/0005055}{{\tt arXiv:quant-ph/0005055
  [quant-ph]}}.

\bibitem{Grinko:2019}
D.~Grinko, J.~Gacon, C.~Zoufal, and S.~Woerner, {\em {Iterative Quantum
  Amplitude Estimation}}.
  \href{http://dx.doi.org/10.1038/s41534-021-00379-1}{npj Quantum Inf {\bf 7}
  (2021) no.~52, }, \href{http://arxiv.org/abs/1912.05559}{{\tt
  arXiv:1912.05559 [quant-ph]}}.

\bibitem{Suzuki:2019}
Y.~Suzuki, S.~Uno, R.~Raymond, T.~Tanaka, T.~Onodera, and N.~Yamamoto, {\em
  {Amplitude estimation without phase estimation}}.
  \href{http://dx.doi.org/10.1007/s11128-019-2565-2}{Quantum Information
  Processing {\bf 19} (2020) no.~75, },
  \href{http://arxiv.org/abs/1904.10246}{{\tt arXiv:1904.10246 [quant-ph]}}.

\bibitem{Nakaji:2020}
K.~Nakaji, {\em {Faster Amplitude Estimation}}.
  \href{http://dx.doi.org/10.26421/QIC20.13-14-2}{Quantum Information \&
  Computation 2020 {\bf 20} (2020) no.~13\&14, },
  \href{http://arxiv.org/abs/2003.02417}{{\tt arXiv:2003.02417 [quant-ph]}}.

\bibitem{Rebentrost2018QuantumCF}
P.~Rebentrost, B.~Gupt, and T.~R. Bromley, {\em Quantum computational finance:
  Monte Carlo pricing of financial derivatives}. Physical Review A {\bf 98}
  (2018)  022321.

\bibitem{Zoufal:2019}
C.~Zoufal, A.~Lucchi, and S.~Woerner, {\em {Quantum Generative Adversarial
  Networks for Learning and Loading Random Distributions}}.
  \href{http://dx.doi.org/10.1038/s41534-019-0223-2}{npj Quantum Inf {\bf 5}
  (2019) no.~103, }, \href{http://arxiv.org/abs/1904.00043}{{\tt
  arXiv:1904.00043 [quant-ph]}}.

\bibitem{Stamatopoulos:2020xez}
N.~Stamatopoulos, D.~J. Egger, Y.~Sun, C.~Zoufal, R.~Iten, N.~Shen, and
  S.~Woerner, {\em {Option Pricing using Quantum Computers}}.
  \href{http://dx.doi.org/10.22331/q-2020-07-06-291}{Quantum {\bf 4} (2020)
  291}, \href{http://arxiv.org/abs/1905.02666}{{\tt arXiv:1905.02666}}.

\bibitem{Stamatopoulos:2021eyd}
N.~Stamatopoulos, G.~Mazzola, S.~Woerner, and W.~J. Zeng, {\em {Towards Quantum
  Advantage in Financial Market Risk using Quantum Gradient Algorithms}}.
  \href{http://arxiv.org/abs/2111.12509}{{\tt arXiv:2111.12509 [quant-ph]}}.

\bibitem{grover2002creating}
L.~Grover and T.~Rudolph, {\em Creating superpositions that correspond to
  efficiently integrable probability distributions}.
  \href{http://arxiv.org/abs/quant-ph/0208112}{{\tt arXiv:quant-ph/0208112
  [quant-ph]}}.

\bibitem{adedoyin2018quantum}
A.~Adedoyin, J.~Ambrosiano, P.~Anisimov, A.~B{\"a}rtschi, W.~Casper,
  G.~Chennupati, C.~Coffrin, H.~Djidjev, D.~Gunter, S.~Karra, {\em et al.},
  {\em Quantum algorithm implementations for beginners}.
  \href{http://arxiv.org/abs/1804.03719}{{\tt arXiv:1804.03719 [cs.ET]}}.

\bibitem{woerner2019quantum}
S.~Woerner and D.~J. Egger, {\em Quantum risk analysis}. npj Quantum
  Information {\bf 5} (2019) no.~1, 1--8,
  \href{http://arxiv.org/abs/1806.06893}{{\tt arXiv:1806.06893 [quant-ph]}}.

\bibitem{gacon2020quantum}
J.~Gacon, C.~Zoufal, and S.~Woerner, ``Quantum-enhanced simulation-based
  optimization,'' in {\em 2020 IEEE International Conference on Quantum
  Computing and Engineering (QCE)}, pp.~47--55, IEEE.
\newblock 2020.
\newblock \href{http://arxiv.org/abs/2005.10780}{{\tt arXiv:2005.10780
  [quant-ph]}}.

\bibitem{holmes2020efficient}
A.~Holmes and A.~Matsuura, ``Efficient quantum circuits for accurate state
  preparation of smooth, differentiable functions,'' in {\em 2020 IEEE
  International Conference on Quantum Computing and Engineering (QCE)},
  pp.~169--179, IEEE.
\newblock 2020.
\newblock \href{http://arxiv.org/abs/2005.04351}{{\tt arXiv:2005.04351
  [quant-ph]}}.

\bibitem{garcia2021quantum}
J.~J. Garc{\'\i}a-Ripoll, {\em Quantum-inspired algorithms for multivariate
  analysis: from interpolation to partial differential equations}. Quantum {\bf
  5} (2021)  431, \href{http://arxiv.org/abs/1909.06619}{{\tt arXiv:1909.06619
  [quant-ph]}}.

\bibitem{Chang:2021ufg}
S.~Y. Chang, S.~Herbert, S.~Vallecorsa, E.~F. Combarro, and R.~Duncan, {\em
  {Dual-Parameterized Quantum Circuit GAN Model in High Energy Physics}}.
  \href{http://dx.doi.org/10.1051/epjconf/202125103050}{EPJ Web Conf. {\bf 251}
  (2021)  03050}, \href{http://arxiv.org/abs/2103.15470}{{\tt arXiv:2103.15470
  [quant-ph]}}.

\bibitem{Kim:2021wrr}
M.~Kim, P.~Ko, J.-h. Park, and M.~Park, {\em {Leveraging Quantum Annealer to
  identify an Event-topology at High Energy Colliders}}.
  \href{http://arxiv.org/abs/2111.07806}{{\tt arXiv:2111.07806 [hep-ph]}}.

\bibitem{Bargassa:2021jmk}
P.~Bargassa, T.~Cabos, S.~Cavinato, A.~Cordeiro Oudot~Choi, and T.~Hessel, {\em
  {Quantum algorithm for the classification of supersymmetric top quark
  events}}. \href{http://dx.doi.org/10.1103/PhysRevD.104.096004}{Phys. Rev. D
  {\bf 104} (2021) no.~9, 096004}, \href{http://arxiv.org/abs/2106.00051}{{\tt
  arXiv:2106.00051 [quant-ph]}}.

\bibitem{Belis:2021zqi}
V.~Belis, S.~Gonz\'alez-Castillo, C.~Reissel, S.~Vallecorsa, E.~F. Combarro,
  G.~Dissertori, and F.~Reiter, {\em {Higgs analysis with quantum
  classifiers}}. \href{http://dx.doi.org/10.1051/epjconf/202125103070}{EPJ Web
  Conf. {\bf 251} (2021)  03070}, \href{http://arxiv.org/abs/2104.07692}{{\tt
  arXiv:2104.07692 [quant-ph]}}.

\bibitem{Heredge:2021vww}
J.~Heredge, C.~Hill, L.~Hollenberg, and M.~Sevior, {\em {Quantum Support Vector
  Machines for Continuum Suppression in B Meson Decays}}.
  \href{http://arxiv.org/abs/2103.12257}{{\tt arXiv:2103.12257 [quant-ph]}}.

\bibitem{Cormier:2019kcq}
K.~Cormier, R.~Di~Sipio, and P.~Wittek, {\em {Unfolding measurement
  distributions via quantum annealing}}.
  \href{http://dx.doi.org/10.1007/JHEP11(2019)128}{JHEP {\bf 11} (2019)  128},
  \href{http://arxiv.org/abs/1908.08519}{{\tt arXiv:1908.08519
  [physics.data-an]}}.

\bibitem{Wei:2019rqy}
A.~Y. Wei, P.~Naik, A.~W. Harrow, and J.~Thaler, {\em {Quantum Algorithms for
  Jet Clustering}}. \href{http://dx.doi.org/10.1103/PhysRevD.101.094015}{Phys.
  Rev. D {\bf 101} (2020) no.~9, 094015},
  \href{http://arxiv.org/abs/1908.08949}{{\tt arXiv:1908.08949 [hep-ph]}}.

\bibitem{Perez-Salinas:2020nem}
A.~P\'erez-Salinas, J.~Cruz-Martinez, A.~A. Alhajri, and S.~Carrazza, {\em
  {Determining the proton content with a quantum computer}}.
  \href{http://dx.doi.org/10.1103/PhysRevD.103.034027}{Phys. Rev. D {\bf 103}
  (2021) no.~3, 034027}, \href{http://arxiv.org/abs/2011.13934}{{\tt
  arXiv:2011.13934 [hep-ph]}}.

\bibitem{Li:2021kcs}
T.~Li, X.~Guo, W.~K. Lai, X.~Liu, E.~Wang, H.~Xing, D.-B. Zhang, and S.-L. Zhu,
  {\em {Partonic Structure by Quantum Computing}}.
  \href{http://arxiv.org/abs/2106.03865}{{\tt arXiv:2106.03865 [hep-ph]}}.

\bibitem{Bepari:2020xqi}
K.~Bepari, S.~Malik, M.~Spannowsky, and S.~Williams, {\em {Towards a quantum
  computing algorithm for helicity amplitudes and parton showers}}.
  \href{http://dx.doi.org/10.1103/PhysRevD.103.076020}{Phys. Rev. D {\bf 103}
  (2021) no.~7, 076020}, \href{http://arxiv.org/abs/2010.00046}{{\tt
  arXiv:2010.00046 [hep-ph]}}.

\bibitem{Ramirez-Uribe:2021ubp}
S.~Ram\'\i{}rez-Uribe, A.~E. Renter\'\i{}a-Olivo, G.~Rodrigo, G.~F.~R.
  Sborlini, and L.~Vale~Silva, {\em {Quantum algorithm for Feynman loop
  integrals}}. \href{http://arxiv.org/abs/2105.08703}{{\tt arXiv:2105.08703
  [hep-ph]}}.

\bibitem{Bauer:2019qxa}
C.~W. Bauer, W.~A. de~Jong, B.~Nachman, and D.~Provasoli, {\em {Quantum
  Algorithm for High Energy Physics Simulations}}.
  \href{http://dx.doi.org/10.1103/PhysRevLett.126.062001}{Phys. Rev. Lett. {\bf
  126} (2021) no.~6, 062001}, \href{http://arxiv.org/abs/1904.03196}{{\tt
  arXiv:1904.03196 [hep-ph]}}.

\bibitem{Williams:2021lvr}
S.~Williams, S.~Malik, M.~Spannowsky, and K.~Bepari, {\em {A quantum walk
  approach to simulating parton showers}}.
  \href{http://arxiv.org/abs/2109.13975}{{\tt arXiv:2109.13975 [hep-ph]}}.

\bibitem{Bravo-Prieto:2021ehz}
C.~Bravo-Prieto, J.~Baglio, M.~C\`e, A.~Francis, D.~M. Grabowska, and
  S.~Carrazza, {\em {Style-based quantum generative adversarial networks for
  Monte Carlo events}}. \href{http://arxiv.org/abs/2110.06933}{{\tt
  arXiv:2110.06933 [quant-ph]}}.

\bibitem{shende2006synthesis}
V.~V. Shende, S.~S. Bullock, and I.~L. Markov, {\em Synthesis of quantum-logic
  circuits}. IEEE Transactions on Computer-Aided Design of Integrated Circuits
  and Systems {\bf 25} (2006) no.~6, 1000--1010,
  \href{http://arxiv.org/abs/quant-ph/0406176}{{\tt arXiv:quant-ph/0406176
  [quant-ph]}}.

\bibitem{byckling1973particle}
E.~Byckling and K.~Kajantie, {\em Particle kinematics}.

\bibitem{Gavin:2013kga}
R.~Gavin, C.~Hangst, M.~Kr\"amer, M.~M\"uhlleitner, M.~Pellen, E.~Popenda, and
  M.~Spira, {\em {Matching Squark Pair Production at NLO with Parton Showers}}.
  \href{http://dx.doi.org/10.1007/JHEP10(2013)187}{JHEP {\bf 10} (2013)  187},
  \href{http://arxiv.org/abs/1305.4061}{{\tt arXiv:1305.4061 [hep-ph]}}.

\bibitem{Gavin:2014yga}
R.~Gavin, C.~Hangst, M.~Kr\"amer, M.~M\"uhlleitner, M.~Pellen, E.~Popenda, and
  M.~Spira, {\em {Squark Production and Decay matched with Parton Showers at
  NLO}}. \href{http://dx.doi.org/10.1140/epjc/s10052-014-3243-2}{Eur. Phys. J.
  C {\bf 75} (2015) no.~1, 29}, \href{http://arxiv.org/abs/1407.7971}{{\tt
  arXiv:1407.7971 [hep-ph]}}.

\bibitem{Cavasonza:2014xra}
L.~Ali~Cavasonza, M.~Kr\"amer, and M.~Pellen, {\em {Electroweak fragmentation
  functions for dark matter annihilation}}.
  \href{http://dx.doi.org/10.1088/1475-7516/2015/02/021}{JCAP {\bf 02} (2015)
  021}, \href{http://arxiv.org/abs/1409.8226}{{\tt arXiv:1409.8226 [hep-ph]}}.

\bibitem{Cavasonza:2016qem}
L.~Ali~Cavasonza, H.~Gast, M.~Kr\"amer, M.~Pellen, and S.~Schael, {\em
  {Constraints on leptophilic dark matter from the AMS-02 experiment}}.
  \href{http://dx.doi.org/10.3847/1538-4357/aa624d}{Astrophys. J. {\bf 839}
  (2017) no.~1, 36}, \href{http://arxiv.org/abs/1612.06634}{{\tt
  arXiv:1612.06634 [hep-ph]}}. [Erratum: Astrophys.J. 869, 89 (2018)].

\bibitem{Lepage:1977sw}
G.~P. Lepage, {\em {A New Algorithm for Adaptive Multidimensional
  Integration}}. \href{http://dx.doi.org/10.1016/0021-9991(78)90004-9}{J.
  Comput. Phys. {\bf 27} (1978)  192}.

\bibitem{Arnold:2008rz}
K.~Arnold {\em et al.}, {\em {VBFNLO: A Parton level Monte Carlo for processes
  with electroweak bosons}}.
  \href{http://dx.doi.org/10.1016/j.cpc.2009.03.006}{Comput. Phys. Commun. {\bf
  180} (2009)  1661--1670}, \href{http://arxiv.org/abs/0811.4559}{{\tt
  arXiv:0811.4559 [hep-ph]}}.

\bibitem{Baglio:2011juf}
J.~Baglio {\em et al.}, {\em {VBFNLO: A Parton Level Monte Carlo for Processes
  with Electroweak Bosons -- Manual for Version 2.7.0}}.
  \href{http://arxiv.org/abs/1107.4038}{{\tt arXiv:1107.4038 [hep-ph]}}.

\bibitem{Baglio:2014uba}
J.~Baglio {\em et al.}, {\em {Release Note - VBFNLO 2.7.0}}.
  \href{http://arxiv.org/abs/1404.3940}{{\tt arXiv:1404.3940 [hep-ph]}}.

\bibitem{Actis:2012qn}
S.~Actis, A.~Denner, L.~Hofer, A.~Scharf, and S.~Uccirati, {\em {Recursive
  generation of one-loop amplitudes in the Standard Model}}.
  \href{http://dx.doi.org/10.1007/JHEP04(2013)037}{JHEP {\bf 04} (2013)  037},
\href{http://arxiv.org/abs/1211.6316}{{\tt arXiv:1211.6316 [hep-ph]}}.

\bibitem{Actis:2016mpe}
S.~Actis, A.~Denner, L.~Hofer, J.-N. Lang, A.~Scharf, and S.~Uccirati, {\em
  {RECOLA: REcursive Computation of One-Loop Amplitudes}}.
  \href{http://dx.doi.org/10.1016/j.cpc.2017.01.004}{Comput. Phys. Commun. {\bf
  214} (2017)  140--173},
\href{http://arxiv.org/abs/1605.01090}{{\tt arXiv:1605.01090 [hep-ph]}}.

\bibitem{Qiskit}
M.~Treinish and et~al., \href{http://dx.doi.org/10.5281/zenodo.5762490}{{\em
  Qiskit/qiskit: Qiskit 0.33.0}}, Dec., 2021.
\newblock \url{https://doi.org/10.5281/zenodo.5762490}.

\bibitem{Denner:2000bj}
A.~Denner, S.~Dittmaier, M.~Roth, and D.~Wackeroth, {\em {Electroweak radiative
  corrections to ${e}^+ {e}^- \to {W W} \to$ 4 fermions in double pole
  approximation: The RACOONWW approach}}.
  \href{http://dx.doi.org/10.1016/S0550-3213(00)00511-3}{Nucl. Phys. {\bf B587}
  (2000)  67--117},
\href{http://arxiv.org/abs/hep-ph/0006307}{{\tt arXiv:hep-ph/0006307
  [hep-ph]}}.

\bibitem{Tanabashi:2018oca}
{\bf ParticleDataGroup} Collaboration, M.~Tanabashi {\em et al.}, {\em {Review
  of Particle Physics}}.
\href{http://dx.doi.org/10.1103/PhysRevD.98.030001}{Phys. Rev. {\bf D98} (2018)
  no.~3, 030001}.

\bibitem{Bardin:1988xt}
D.~{\relax Yu}. Bardin, A.~Leike, T.~Riemann, and M.~Sachwitz, {\em
  {Energy-dependent width effects in ${e}^+ {e}^-$-annihilation near the
  Z-boson pole}}.
\href{http://dx.doi.org/10.1016/0370-2693(88)91627-9}{Phys. Lett. {\bf B206}
  (1988)  539--542}.

\bibitem{qram}
V.~Giovannetti, S.~Lloyd, and L.~Maccone, {\em {Quantum Random Access Memory}}.
  \href{http://dx.doi.org/10.1103/PhysRevLett.100.160501}{Phys. Rev. Lett. {\bf
  100} (2008)  160501}.

\bibitem{schuld2021}
M.~Schuld, R.~Sweke, and J.~J. Meyer, {\em {Effect of data encoding on the
  expressive power of variational quantum-machine-learning models}}.
  \href{http://dx.doi.org/10.1103/PhysRevA.103.032430}{Phys. Rev. A {\bf 103}
  (2021)  }.

\bibitem{goodfellow2014generative}
I.~Goodfellow, J.~Pouget-Abadie, M.~Mirza, B.~Xu, D.~Warde-Farley, S.~Ozair,
  A.~Courville, and Y.~Bengio, {\em Generative adversarial nets}. Advances in
  neural information processing systems {\bf 27} (2014)  .

\bibitem{gui2021review}
J.~Gui, Z.~Sun, Y.~Wen, D.~Tao, and J.~Ye, {\em A review on generative
  adversarial networks: Algorithms, theory, and applications}. IEEE
  Transactions on Knowledge and Data Engineering (2021)  ,
  \href{http://arxiv.org/abs/2001.06937}{{\tt arXiv:2001.06937 [cs]}}.

\bibitem{variatalgo2021}
M.~Cerezo, A.~Arrasmith, R.~Babbush, S.~C. Benjamin, S.~Endo, K.~Fujii, J.~R.
  McClean, K.~Mitarai, X.~Yuan, L.~Cincio, and et~al., {\em {Variational
  quantum algorithms}}.
  \href{http://dx.doi.org/10.1038/s42254-021-00348-9}{Nature Reviews Physics
  {\bf 3} (2021) no.~9, }.

\bibitem{agliardi2022optimal}
G.~Agliardi and E.~Prati, {\em Optimal tuning of quantum generative adversarial
  networks for multivariate distribution loading}. Quantum Reports {\bf 4}
  (2022) no.~1, 75--105.

\bibitem{nielsen_chuang_2010}
M.~A. Nielsen and I.~L. Chuang,
  \href{http://dx.doi.org/10.1017/CBO9780511976667}{{\em Quantum Computation
  and Quantum Information: 10th Anniversary Edition}}.
\newblock Cambridge University Press, 2010.

\bibitem{shor1999polynomial}
P.~W. Shor, {\em Polynomial-time algorithms for prime factorization and
  discrete logarithms on a quantum computer}. SIAM review {\bf 41} (1999)
  no.~2, 303--332.

\bibitem{moll2018quantum}
N.~Moll, P.~Barkoutsos, L.~S. Bishop, J.~M. Chow, A.~Cross, D.~J. Egger,
  S.~Filipp, A.~Fuhrer, J.~M. Gambetta, M.~Ganzhorn, {\em et al.}, {\em Quantum
  optimization using variational algorithms on near-term quantum devices}.
  Quantum Science and Technology {\bf 3} (2018) no.~3, 030503,
  \href{http://arxiv.org/abs/1710.01022}{{\tt arXiv:1710.01022 [quant-ph]}}.

\bibitem{preskill_quantum_2018}
J.~Preskill, {\em Quantum Computing in the {NISQ} era and beyond}.
  \href{http://arxiv.org/abs/1801.00862}{{\tt 1801.00862}}.
  \url{http://arxiv.org/abs/1801.00862}.

\end{thebibliography}\endgroup
\end{document}